\documentclass[cameraready]{Interspeech}
\usepackage{amsmath,graphicx,hyperref, multicol, multirow, tabularx, booktabs, subcaption, makecell, amssymb, booktabs}
\title{From Masking to Merging:\\ Rethinking SpecAugment for Efficient Audio Spectrogram Transformer}

\author[orcid=0009-0008-5869-5892]{Minhee}{Park}
\author[orcid=0009-0008-7915-3056]{Hyowon}{Ahn}
\author[orcid=0000-0003-4085-2470, correspondingauthor]{Chanwoo}{Kim}
\address{
    Korea University, Republic of Korea
}

\email{himinsunsine@gmail.com, \{hw1204, chanwcom\}@korea.ac.kr}

\keywords{Audio Augmentation, Audio Classification, Token Reduction, Efficiency}

\usepackage{comment}
\usepackage{url}
\hypersetup{hidelinks}
\begin{document}

\maketitle

\begin{abstract}
This paper proposes SpecAugment-Patch Merging, a simple yet effective method to accelerate Audio Spectrogram Transformer (AST) training.  
We first apply SpecAugment to mask input spectrograms at the patch level, and after positional embeddings are added, the method selects r pairs of masked patches and merges them, reducing the number of tokens processed by the Transformer.  
Increasing the number of merged pairs r from 0 to 100 keeps mAP on AudioSet nearly unchanged (34.07 to 34.08) while throughput increases from 43.3 to 49.3 samples/sec, which is a relatively 13.9\% improvement. Similar patterns appear on ESC-50 and Speech Commands V2, where throughput steadily improves with only minor accuracy changes, demonstrating that this merging approach provides faster training with minimal performance loss.
\end{abstract}

\section{Introduction}
\label{sec:intro}
Recent deep learning models for audio understanding, most notably the Audio
Spectrogram Transformer (AST) \cite{gong2021ast}, have achieved remarkable
performance across a wide range of benchmarks. AST operates directly on 2D audio
spectrogram patches, making strong data augmentation particularly important to
prevent overfitting and improve generalization. A variety of data augmentation
techniques have been proposed in speech processing to improve robustness against
mismatched acoustic conditions and to enhance model regularization
\cite{x_cui_taslp_2015_00, n_jaitly_icml_workshop_2013_00,
c_kim_interspeech_2019_00}.

Among the various augmentation
strategies applied to AST, SpecAugment \cite{park19e_interspeech} has become
a standard technique. It enhances robustness by randomly applying time mask and
frequency mask to the input spectrogram, thereby forcing the model to learn
contextually rich representations.
However, SpecAugment inherently produces masked regions with limited spectral information. Although these regions contribute little semantic content, their embedding tokens are still processed by the transformer blocks, resulting in unnecessary computational overhead. Thus, while SpecAugment enhances generalization, its impact on computational efficiency has not been explicitly addressed. 

From a computational perspective, transformer-based models  incur quadratic
complexity with respect to the number of input tokens due to self-attention
\cite{a_vaswani_nips_2017_00}. To
mitigate this cost, prior work has explored both attention optimization
techniques \cite{dao2022flashattention, dao2024flashattention} and
token-reduction strategies \cite{koutini2022efficient, behera24_interspeech,
bolyatoken}. While these approaches improve efficiency from architectural or
algorithmic perspectives, token-reduction methods often require additional
modules or similarity computations to determine merge candidates, or rely on
random dropping without explicitly considering the information content of
individual tokens. In contrast, our approach leverages augmentation-induced
masked regions as inherent structural cues for token reduction, avoiding
additional architectural complexity.

Based on this insight, we propose SpecAugment-Patch Merging. The maximum mask settings of AST’s original SpecAugment \cite{park19e_interspeech} are converted into patch units (16×16, stride 10) to record mask positions, and a subset of the masked patches is randomly merged, as described in Section \ref{sec:METHOD}.
Merging occurs after patch and positional embeddings but just before the transformer encoder.

\begin{figure}
    \centering 
    \includegraphics[width=\linewidth]{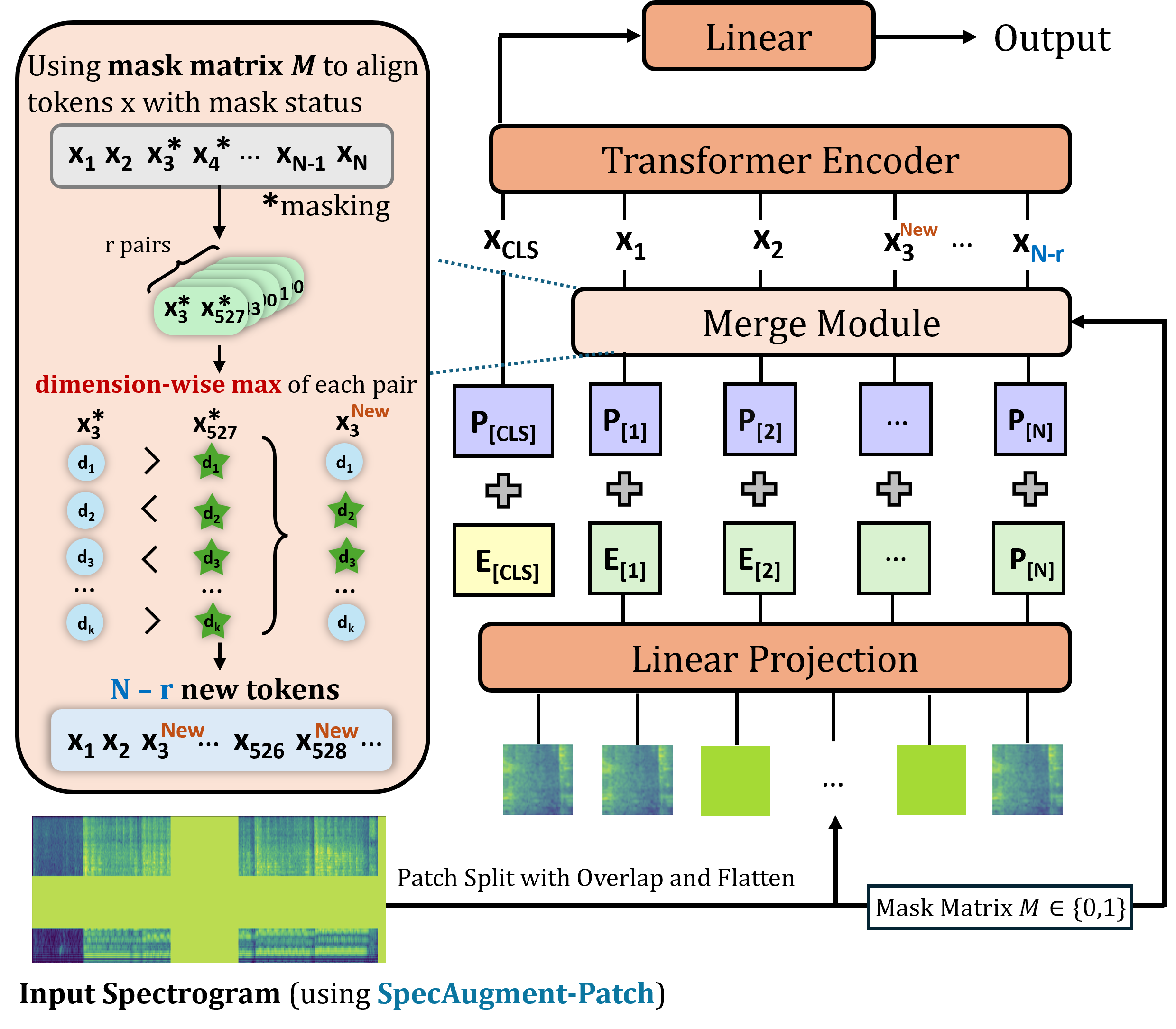}
    \caption{Overall architecture of the proposed SpecAugment-Patch Merging AST.}
    \label{fig:architecture}
\end{figure}

 We evaluate the proposed
method on large-scale audio datasets and confirm that SpecAugment-Patch Merging
improves processing speed while maintaining accuracy 
with only minimal loss even under strong augmentation.\footnote{The source code 
is publicly available at
\url{https://github.com/slp-lab-research/specaug_patch_merge}.}

The main contributions are:
\begin{itemize}
\item Viewing masked patches from data augmentation as opportunities for token merging rather than mere information loss.
\item We comprehensively evaluate performance retention and efficiency improvements on diverse audio classification datasets and verify the effectiveness of the proposed approach through additional experiments.
\item Demonstrating that, under identical token reduction ratios to PaSST-U \cite{koutini2022efficient}, our method achieves higher training throughput while maintaining competitive accuracy, highlighting the efficiency advantage of augmentation-aware token merging.
\end{itemize}

These results show that the redundancy and inactive regions introduced by augmentation can be transformed into efficient learning signals.

\section{Related Work}
\label{sec:Related Work}

\subsection{Audio Augmentation}

Various audio augmentation methods improve classification and recognition.
Among them, SpecAugment \cite{park19e_interspeech} is a widely used method in speech recognition, applying time warping, frequency masking, and time masking directly to the spectrogram. In frequency masking, a start position $s_f$ and width $w_f$ are selected to zero out consecutive Mel channels; similarly, in time masking, a start $s_t$ and width $w_t$ are chosen to mask consecutive time steps. The mask widths are uniformly sampled up to preset limits, and the start positions are uniformly sampled within the valid range.

Mixup \cite{zhang2018mixup} creates virtual samples by linearly blending two
inputs and their labels with a mixing ratio $\lambda$, and shows strong gains
when combined with SpecAugment in AST. Other methods such as time stretching,
pitch shifting, noise injection, and reverberation
\cite{c_kim_interspeech_2017_00, c_kim_interspeech_2018_00, 7953152} further enhance robustness.

\subsection{Token Reduction for Efficient Audio Transformers}

Transformer-based audio models convert a spectrogram into patch tokens, and the number of tokens directly drives computation and memory use. To efficiently handle large-scale datasets, various token-reduction techniques have been proposed to improve computational efficiency. 

PaSST \cite{koutini2022efficient} applies Patchout, which randomly drops a subset of patches after patch extraction, projection, and positional encoding but before the transformer encoder.
This reduces input tokens to lower computational and memory cost and enables simple cropping of time encodings for fine-tuning or inference on shorter clips. In particular, PaSST-U adopts a patch-level dropping strategy, where tokens are reduced at the patch level. In this sense, it can be regarded as a patch-based token reduction approach, conceptually similar to our method in that both operate on patch-level units for reducing the number of transformer input tokens.

Token Merging (ToMe) \cite{bolyatoken} hierarchically merges pairs of tokens with similar representations to progressively reduce the token set during processing. Originally proposed for ViT \cite{dosovitskiy2021an}, it has been extended to audio in FastAST \cite{behera24_interspeech}, which controls the degree of merging to cut computation and memory on large-scale datasets such as AudioSet \cite{7952261}.

\subsection{Relation to Our Work}

Recent advances in efficient audio transformers have demonstrated that token reduction can substantially reduce computational cost while preserving accuracy. Different from these methods, our approach integrates SpecAugment and token merging within a unified framework, converting augmentation-induced redundancy into computational efficiency.

In Section~\ref{ssec:Results}, we compare our method primarily with PaSST-U, as the implementation of FastAST is not publicly available, preventing a controlled and reproducible comparison. PaSST provides an established and reproducible baseline for evaluating efficiency under unified training settings.

\section{SpecAugment-Patch Merging}
\label{sec:METHOD}
This section presents our method’s three components, which are the architecture, SpecAugment-Patch, and merging, with the last two serving as the core algorithm for efficiency and regularization.

\subsection{Architecture}

The overall architecture is based on the Audio Spectrogram Transformer (AST) \cite{gong2021ast}.
The input audio is first converted into a 2D audio spectrogram, after which SpecAugment-Patch is applied to randomly mask regions along the time and frequency bands.
The masked spectrogram is then divided into fixed-size patches, and each patch is processed with patch embedding and positional embedding to form a sequence of tokens $x$.
Merging is performed after the positional embeddings are added, that is, right before the tokens are fed into the Transformer encoder blocks.
This design ensures that the tokens being merged already contain positional information, allowing the remaining tokens after merging to partially preserve time- and frequency-position cues that the encoder can interpret. The resulting tokens are then processed by the Transformer encoder, and the final representation is used for audio classification.
The detailed algorithm is described in Sections~\ref{ssec:SpecAugment-Patch} and~\ref{ssec: Merging}. The overall architecture is illustrated in Figure~\ref{fig:architecture}.

\begin{table}[ht]
\centering
\caption{Results (mAP) and training throughput S/s (samples/sec) comparison between SpecAugment and SpecAugment-Patch on Balanced AudioSet, both combined with Mixup, at $r=0$.}
\begin{tabular}{ccc}
\Xhline{1.1pt}
Method        & mAP     &S/s  \\ \Xhline{1.1pt}
SpecAugment \cite{park19e_interspeech}      & 34.11 $\pm$ 0.35 & 43.3 \\
SpecAugment-Patch & 34.07 $\pm$ 0.18 & 43.3 \\ \Xhline{1.1pt}
\end{tabular}

\label{tab:specaug}
\end{table}

\begin{table}[ht]
\centering
\caption{Comparison of token-reduction strategies on Balanced AudioSet (merged pairs $r=90$).
Dimension-wise max, mean, and sum merging are compared with random drop.}
\resizebox{\linewidth}{!}{%
\begin{tabular}{cccccc}
\Xhline{1.1pt}
\multirow{2}{*}{}  & \multirow{2}{*}{Drop} & \multicolumn{3}{c}{Merge} \\ \cline{3-5}
                   &                       & Max & Sum & Mean \\ \Xhline{1.1pt}
mAP & 34.03 $\pm$0.11 & \textbf{34.10 $\pm$ 0.31} & 34.07 & 34.00 \\
S/s & 48.5 S/s & 48.63 S/s & 48.6 S/s & 48.9 S/s \\ \Xhline{1.1pt}
\end{tabular}
}
\label{tab:ablation}
\end{table}

\subsection{SpecAugment-Patch}
\label{ssec:SpecAugment-Patch}
SpecAugment-Patch follows the masking principle of the original SpecAugment \cite{park19e_interspeech} used in AST \cite{gong2021ast} but applies masking to the input spectrogram
\(\mathbf{X}\in\mathbb{R}^{T\times F}\) before the patch-embedding stage, aligning the masks to a \(16\times16\) patch grid.
Just as the original SpecAugment directly specifies the maximum time mask width and maximum frequency mask width on the raw spectrogram,
our method similarly pre-defines the maximum numbers of consecutive patches along the time axis \((k_t^{\max})\)
and the frequency axis \((k_f^{\max})\) to control the masking intensity, where the goal is to ensure that the patch-level masks cover at least the same maximum time and frequency spans as the original SpecAugment.

\begin{figure}[t]
    \centering
    \begin{subfigure}[b]{0.48\linewidth}
        \centering 
        \includegraphics[width=\linewidth]{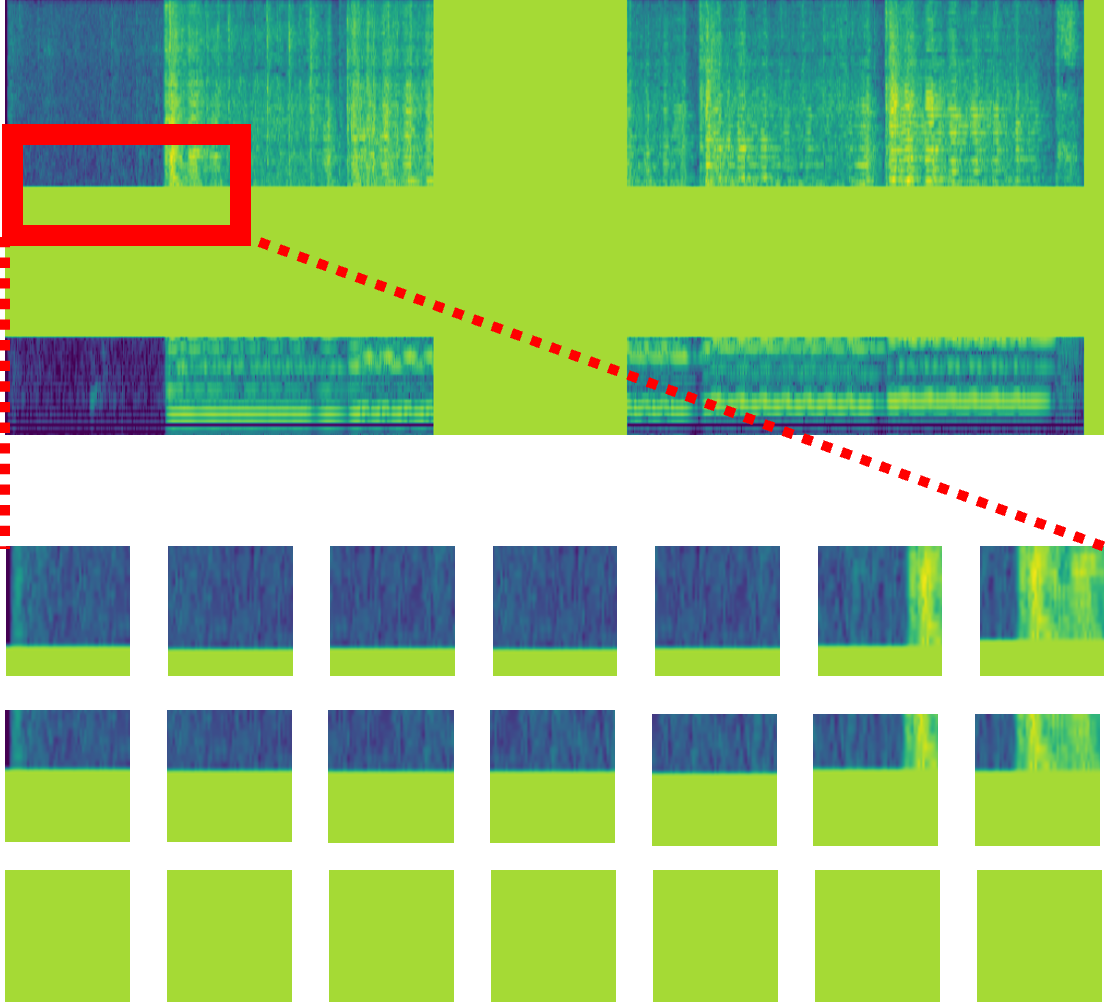}\\[2pt]
        \caption{SpecAugment}
        \label{fig:architecture_a}
    \end{subfigure}
    \hfill
    \begin{subfigure}[b]{0.48\linewidth}
        \centering
        \includegraphics[width=\linewidth]{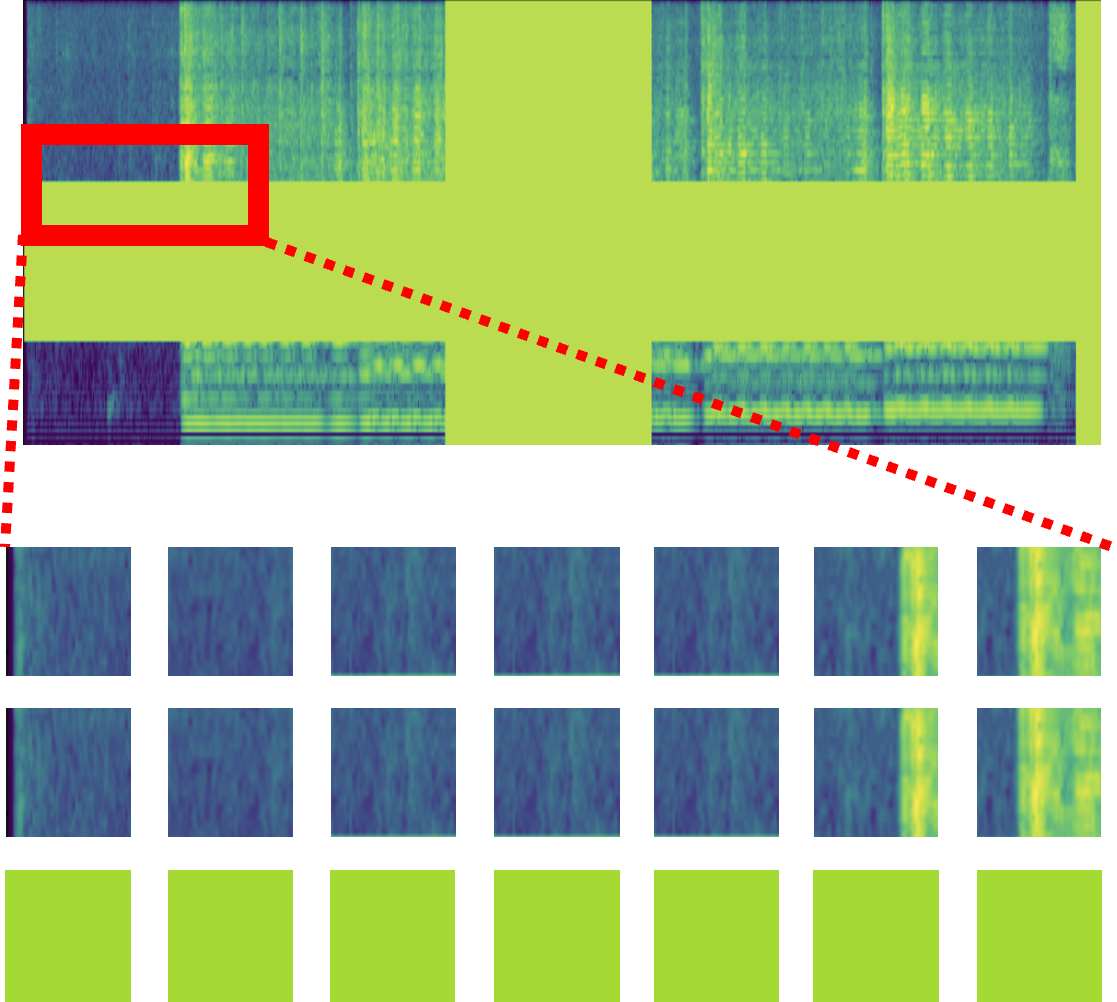}\\[2pt]
        \caption{SpecAugment-Patch}
        \label{fig:architecture_b}
    \end{subfigure}
    \caption{Comparison of spectrogram masking.
(a) SpecAugment masks continuous frequency/time frames, so when the spectrogram is later split into 16×16 patches (stride = 10), a patch can contain both masked and unmasked regions.
(b) SpecAugment-Patch masks along the patch grid, ensuring each patch is either fully masked or fully retained.
The bottom row shows the 16×16 patches from the red-boxed area above.
}
    \label{fig:specaugment}
\end{figure}

Given a patch size \(p\) and stride \(s\), the starting indices along the time and frequency axes are computed to enumerate all valid patch positions,
and the numbers of patches along each axis are
\begin{equation}
N_t=\Big\lfloor \frac{T-p}{s}\Big\rfloor + 1,\qquad 
N_f=\Big\lfloor \frac{F-p}{s}\Big\rfloor + 1 .
\label{eq:patch_num}
\end{equation}
The actual numbers of consecutive patches from the random positions $s_t, s_f$ to be masked are then drawn at random,
\begin{equation}
k_t \sim \mathrm{Uniform}\{1,\dots,k_t^{\max}\},
k_f \sim \mathrm{Uniform}\{1,\dots,k_f^{\max}\},
\end{equation}
and the corresponding mask spans are calculated as
\begin{equation}
w_t = p + (k_t-1)s,\qquad
w_f = p + (k_f-1)s ,
\end{equation}
after which the selected time-frequency regions of the spectrogram are filled with zeros (see Figure~\ref{fig:specaugment}). 

For example, applying Equation~\ref{eq:patch_num} to the AudioSet \cite{7952261} configuration ($T=1024, F=128, p=16, s=10$) gives $N_t = 101$ and $N_f=12$.
To match the original SpecAugment setting used in AST, which applies a maximum time mask width of 192 frames, $ k_t^{\max}\approx 19,
w_t \approx 16 + (19-1)\times10 = 196\ \text{frames}, $
and to match the maximum frequency mask width of 48 bins, $
k_f^{\max}\approx 5,
w_f \approx 16 + (5-1)\times10 = 56\ \text{bins}$.

In other words, the maximum masked patch counts are set to satisfy (or slightly exceed) the original mask widths so that a comparable masking strength is achieved.

After masking, the entire spectrogram is scanned again patch by patch to check whether each patch region
\(\mathbf{X}[t:t+p,\ f:f+p]\) is completely zeroed.
This yields a binary mask matrix \(\mathbf{M}\in\{0,1\}^{N_t\times N_f}\),
which is flattened and later utilized at the token-merging stage.
The zeroed patches remain part of the token sequence after patch embedding and serve as merge candidates. Importantly, when the number of merging pairs $r$ is set to 0, the proposed formulation produces nearly identical mAP and training throughput to the original SpecAugment, as shown in Table~\ref{tab:specaug}, confirming that the augmentation effect is preserved.

\subsection{Merging}
\label{ssec: Merging}
The merging process is performed after adding positional embeddings and right before the tokens are fed into the Transformer encoder.
Let $\mathbf{M}\in\{0,1\}^{N_t\times N_f}$ be the binary mask generated in the masking stage,
where $\mathbf{M}_{t,f}=1$ indicates a fully zeroed patch.
We flatten $\mathbf{M}$ into a candidate vector
$\mathbf{c}\in\{0,1\}^{P}$ ($P = N-2$, excluding the two special tokens ([CLS], [Dist])).
Only positions with $\mathbf{c}_p=1$ are eligible for merging.

For each sample, we select \textbf{$2r$} distinct candidate patches uniformly at random and randomly permute them to form \textbf{$r$} disjoint pairs.
For every pair $(i,j)$, the two token embeddings are compared dimension by dimension,
and a dimension-wise max operation is applied to produce a merged token:
\begin{equation}
\mathbf{z}=\max(\mathbf{x}_i,\mathbf{x}_j)\quad \text{(dimension-wise)}.
\end{equation}
The merged token $\mathbf{z}$ overwrites position $i$, and position $j$ is marked for removal.
After processing all $r$ pairs, the sequence is compacted by discarding the removed positions,
yielding an output of length $N-r$ (the two special tokens are preserved at the front).

A comparison of merging strategies (Table \ref{tab:ablation}) evaluates dimension-wise max, mean, sum, and random drop. Although the performance differences among the merging strategies are relatively small, we selected dimension-wise max merging based on empirical performance and stability. This merging strategy is consistently applied across all datasets in our experiments.

\section{Experiments}
\label{sec:majhead}

This section first describes the datasets and training settings and then presents the performance changes and throughput S/s (samples/sec) as a function of the number of merged pairs $r$.

\subsection{Datasets}
Experiments are conducted on 3 widely used benchmark datasets for audio classification: AudioSet \cite{7952261}, ESC-50 \cite{10.1145/2733373.2806390}, and Speech Commands V2  \cite{warden2018speechcommandsdatasetlimitedvocabulary}. AudioSet is a large-scale, weakly labeled dataset with 527 sound event classes, where each audio clip is up to 10 seconds long, and it is commonly used for multi-label classification. The dataset includes a full training set of over two million clips and a balanced training set of about 22k clips, and a test set of about 20k clips; we use the balanced training set. ESC-50 is a benchmark for environmental sound classification, consisting of 2,000 clips across 50 classes, with each clip fixed at 5 seconds, and is used for single-label classification. Speech Commands V2 is a benchmark for keyword spotting, containing 105,829 one-second recordings from 35 categories sampled at 16 kHz. The training, validation, and test sets contain 84,843, 9,981, and 11,005 samples, respectively, and this dataset is also used for single-label classification.
\label{ssec:Datasets}
\begin{table*}[t]
\centering
\caption{Results on Balanced AudioSet. We report mean ±  standard deviation of mAP, training throughput S/s (samples/sec), and peak GPU memory (GB). Due to different numbers of patch tokens in AST and PaSST-U, we match the merging rate (\%) and report the corresponding number of merged pairs r. Memory reductions relative to the 0\% merging baseline are shown in parentheses.}
\begin{tabular}{ccccccccc}
\Xhline{1.1pt}
\\[-9pt]
\multicolumn{4}{c}{\textbf{AST} (1212 Patches)}                        & \multirow{2}{*}{\textbf{Merge rate(\%)}} & \multicolumn{4}{c}{\textbf{PaSST-U} (1190 Patches)}           \\ \cline{1-4} \cline{6-9} \\[-9pt]
mAP              & S/s  & Mem (GB) & \textbf{r(pairs)} &                                & \textbf{r(pairs)} & mAP              & S/s  & Mem (GB) \\ \Xhline{1.1pt} \\[-9pt]
34.07 $\pm$ 0.18 & 43.3 &  24.64 ( 0.00 GB)  & \textbf{0}     & \textbf{0\%}   & \textbf{0}    & 29.43 $\pm$ 0.14 & 42.7 &  24.13 ( 0.00 GB) \\
34.12 $\pm$ 0.21 & 45.3 &  22.33 ( -2.31 GB)  & \textbf{30}    & \textbf{5\%}   & \textbf{29}   & 29.47 $\pm$ 0.15 & 43.3 &  23.11 ( -1.02 GB) \\
34.04 $\pm$ 0.38 & 46.2 &  22.79 ( -1.85 GB)  & \textbf{60}    & \textbf{10\%}  & \textbf{59}   & 29.62 $\pm$ 0.06 & 44.5 &  22.31 ( -1.82 GB) \\
34.10 $\pm$ 0.31 & 48.6 &  21.85 ( -2.79 GB)  & \textbf{90}    & \textbf{15\%}  & \textbf{88}   & 29.27 $\pm$ 0.21 & 46.2 &  21.29 ( -2.84 GB) \\
34.08 $\pm$ 0.24 & 49.3 &  21.50 ( -3.14 GB)  & \textbf{100}   & \textbf{16.5\%}& \textbf{98}   & 29.58 $\pm$ 0.27 & 46.7 &  21.00 ( -3.13 GB) \\ 
\Xhline{1.1pt}
\end{tabular}

\label{tab:balanced_audioset}
\end{table*}

\begin{table}[ht]
\centering
\caption{Results on ESC-50 and Speech Commands V2.}
\begin{tabularx}{\columnwidth}{@{}c c *{2}{>{\centering\arraybackslash}X}@{}}
\Xhline{1.1pt}
Dataset & r & Acc & S/s \\ \Xhline{1.1pt}
\multirow{6}{*}{ESC-50}
 & 0   & 89.20 $\pm$ 0.43 & 127.3 \\
 & 10  & 88.96 $\pm$ 0.45 & 129.3 \\
 & 20  & 89.00 $\pm$ 0.62 & 130.8 \\
 & 30  & 88.96 $\pm$ 0.23 & 134.6 \\
 & 40  & 88.80 $\pm$ 0.17 & 138.9 \\
 & 50  & 88.67 $\pm$ 0.21 & 142.9 \\ \hline
\multirow{4}{*}{\makecell{Speech\\Commands V2}}
 & 0   & 98.13 $\pm$ 0.01 & 411.1 \\
 & 5  & 98.17 $\pm$ 0.07 & 413.1 \\
 & 10  & 98.14 $\pm$ 0.02 & 420.8 \\
 & 15  & 98.06 $\pm$ 0.04 & 429.4 \\
 \Xhline{1.1pt}
\end{tabularx}

\label{tab:combined}
\end{table}
\subsection{Training Details}
\label{ssec:details}
\subsubsection{Training Settings}

All experiments are initialized from an ImageNet-pretrained DeiT-Base distilled backbone (87M parameters) \cite{dosovitskiy2021an}. Experiments are conducted on a single NVIDIA GeForce RTX 4090 GPU. Mixup \cite{zhang2018mixup} is applied for AudioSet and Speech Commands V2 with ratios of 0.5 and 0.6, respectively, while ESC-50 is trained without mixup. SpecAugment-Patch is applied to all datasets.

Balanced AudioSet is trained for 25 epochs with a learning rate of $5\times10^{-5}$. The learning rate scheduler starts at epoch 10 and decays by a factor of 0.5 every 5 epochs. Weight averaging is applied from epochs 6 to 25.
ESC-50 is trained for 25 epochs with a learning rate of $1\times10^{-4}$. A learning rate scheduler starting at epoch 5 decays the rate by 0.85 at every epoch. 
Speech Commands V2 is trained for 30 epochs with a learning rate of $2.5\times10^{-4}$, using the same scheduler configuration as ESC-50.
Performance is evaluated using mean Average Precision (mAP) for AudioSet and classification accuracy (Acc) for ESC-50 and Speech Commands V2. Training throughput S/s (samples/sec) is measured based on wall-clock time per iteration over the full training step.

\subsubsection{Dataset-Specific Token Configuration}

Balanced AudioSet contains 1212 patches ($101\times12$), ESC-50 contains about 600 patches ($50\times12$), and Speech Commands V2 contains approximately 144 patches ($12\times12$). Since merge candidates are restricted to masked patches, the maximum allowable merge ratio depends on the number of fully masked patches in each dataset.

To ensure sufficient merge candidates, the minimum frequency-mask width is set to two rows. This yields approximately 212, 110, and 34 masked patches for AudioSet, ESC-50, and Speech Commands V2, respectively. The merge parameter $r$ is selected such that $2r$ does not exceed these minimum counts.

\subsection{Results}
\label{ssec:Results}
\subsubsection{Throughput vs. Accuracy Trade-off}
As shown in Table~\ref{tab:balanced_audioset}, on Balanced AudioSet, increasing the merge parameter $r$ from 0 to 100 maintains nearly identical mAP (34.07 $\pm$ 0.18 → 34.08 $\pm$ 0.24) while significantly improving training throughput from 43.3 to 49.3 samples/sec. Memory usage is also reduced from 24.64 GB to 21.50 GB. These results indicate that merging masked patches effectively reduces the number of processed tokens without degrading performance beyond statistical variation.

Similar trends are observed on ESC-50 and Speech Commands V2, as summarized in Table~\ref{tab:combined}. For ESC-50, accuracy slightly decreases from 89.20 $\pm$ 0.43 to 88.67 $\pm$ 0.21 as $r$ increases, while throughput steadily rises from 127.3 to 142.9 S/s. For Speech Commands V2, accuracy remains highly stable (98.13 $\pm$ 0.01 → 98.06 $\pm$ 0.04), with throughput improving from 411.1 to 429.4 S/s. Overall, the proposed merging strategy achieves a favorable trade-off between computational efficiency and predictive performance.
\subsubsection{Comparison with PaSST-U}
For fair comparison, PaSST-U is retrained under the same optimizer, scheduler, learning rate, batch size, number of epochs, and input sampling rate (16 kHz) as AST. The PaSST paper used 32 kHz audio and was trained on full AudioSet; here, we unify the sampling rate and dataset setting to ensure a controlled comparison. The quantitative comparison is reported in Table~\ref{tab:balanced_audioset}. Under matched merge ratios, SpecAugment-Patch Merging on AST consistently achieves higher training throughput than PaSST-U while maintaining competitive accuracy. For example, at 16.5\% merge rate, AST reaches 49.3 S/s compared to 46.7 S/s for PaSST-U. Memory reduction trends are also comparable, with both methods achieving over 3 GB reduction relative to the no-merge baseline.

Due to differences in implementation details and training configurations, absolute mAP values may differ from those reported in the PaSST paper. Therefore, the comparison focuses primarily on efficiency improvements under a unified training setup.

\section{Conclusion}
\label{sec:CONCLUSION}
This work proposes SpecAugment-Patch Merging, which treats masked spectrogram regions as opportunities for token reduction. By converting masks into patch units and merging a subset, the method reduces transformer tokens and achieves faster training with minimal loss across multiple audio benchmarks. The results demonstrate consistent improvements in training throughput while maintaining competitive classification performance, indicating that the proposed approach offers an effective balance between computational efficiency and predictive accuracy. These findings suggest that leveraging augmentation-induced sparsity can serve as a practical strategy for improving the efficiency of transformer-based audio models.

While leveraging augmentation-induced masked regions provides a simple and principled mechanism for token reduction, the restriction to these regions inherently limits the maximum achievable reduction ratio. Nevertheless, the proposed strategy can be readily extended to other patch-based transformer architectures that employ SpecAugment, suggesting broader applicability beyond AST.
\section{Acknowledgments}
This work was supported in part by: the Institute of Information \& Communications Technology Planning \& Evaluation (IITP) grant funded by the Korean government (MSIT) under Grant No. RS-2019-II190079 for the Artificial Intelligence Graduate School Program at Korea University; the Institute of Information \& Communications Technology Planning \& Evaluation (IITP) grant funded by the Korean government (MSIT) under Grant No. RS-2025-02304828 for the Artificial Intelligence Star Fellowship Support Program to Nurture the Best Talents; the Institute of Information \& Communications Technology Planning \& Evaluation (IITP) grant funded by the Korean government (MSIT) under Grant No. RS-2025-25442867; the National Research Foundation of Korea (NRF) grant funded by the Korean government (MSIT) under Grant No. RS-2025-24535409; the Technology Development Program funded by the Ministry of SMEs and Startups (MSS, Korea) under Grant No. RS-2026-25534256; and the Supreme Prosecutor's Office Research Grant in 2026 (research title: Development of fake voice detection technology robust in new voice generation technology and speaker recognition).
\section{Use of Generative AI Disclosure}
Generative AI tools were used only for language editing and polishing of the manuscript.

\bibliographystyle{IEEEtran}
\bibliography{common_bib_file}

\end{document}